\documentclass[aps,prl,twocolumn,showpacs,superscriptaddress]{revtex4-2}

\usepackage{graphicx}
\usepackage{dcolumn}
\usepackage{bm}

\usepackage[utf8]{inputenc}
\usepackage[T1]{fontenc}
\usepackage{mathptmx}
\usepackage{etoolbox}
\usepackage{amsfonts}
\usepackage{amsmath}
\usepackage{physics}
\usepackage{enumerate}
\usepackage{amssymb}
\usepackage{enumitem}

\begin{document}

\title{Geometric Aspects of Entanglement}


\author{Lucio De Simone}
\email{l.desimone3@student.unisi.it}
\affiliation{DSFTA, University of Siena, Via Roma 56, 53100 Siena, Italy}
\affiliation{INFN Sezione di Perugia, 06123 Perugia, Italy}

\author{Lorenzo Capra}
\email{lorenzo.capra@student.unisi.it}
\affiliation{DSFTA, University of Siena, Via Roma 56, 53100 Siena, Italy}
\affiliation{INFN Sezione di Perugia, 06123 Perugia, Italy}

\author{Arthur Vesperini}
\email{arthur.vesperini@unisi.it}
\affiliation{ Wilhelm-Johnen-Straße, 52428 Jülich, Germany}

\author{Leonardo Rossi}
\email{l.rossi59@student.unisi.it}
\affiliation{DSFTA, University of Siena, Via Roma 56, 53100 Siena, Italy}

\author{Loris Di Cairano}
\email{loris.dicairano@uni.lu}
\affiliation{Department of Physics and Materials Science, University of Luxembourg, Luxembourg City, L-1511, Luxembourg}

\author{Roberto Franzosi}
\email{roberto.franzosi@unisi.it}
\affiliation{DSFTA, University of Siena, Via Roma 56, 53100 Siena, Italy}
\affiliation{INFN Sezione di Perugia, 06123 Perugia, Italy}

\date{\today}


\begin{abstract}
Quantum entanglement is a fundamental resource in quantum information theory, yet its general characterization and quantification remain challenging, especially in multipartite systems. In this work we investigate entanglement from a geometric perspective, focusing on the Riemannian structure induced by the Fubini--Study metric on the projective Hilbert space of multi-qubit quantum states. By exploiting the local-unitary invariance of this metric, we derive the entanglement distance (ED), a geometric measure that quantifies entanglement as an obstruction to locally minimizing the sum of squared Fubini--Study distances generated by local operations.

We analyze the properties of ED for pure multi-qubit states and discuss its behavior under local operations and classical communication. In particular, we show that ED reproduces established entanglement measures in well-defined and restricted settings. For pure states of two qubits, ED reduces to an exact monotone function of the concurrence and, independently, to an explicit monotone function of the entropy of entanglement. These results provide a clear geometric interpretation of standard bipartite entanglement measures within the present framework, while highlighting the limitations of such correspondences beyond the two-qubit case.
\end{abstract}

\maketitle




\section{Introduction}

Entanglement plays a central role in quantum information theory and underpins many of its applications in emerging quantum technologies.
It is widely recognized as a fundamental resource in quantum cryptography, teleportation, quantum computation, and quantum metrology applications \cite{GUHNE20091, alireza, PhysRevA.105.052439}.

Despite its importance, entanglement remains a conceptually elusive phenomenon. Its general characterization and quantification still pose significant open challenges, particularly in systems beyond the bipartite case \cite{PhysRevA.95.062116, PhysRevA.67.022320,ScienticReports1-13}.
Over the past decades, a substantial body of literature has addressed the problem of entanglement quantification. However, rigorous and well-established results have largely been confined to bipartite systems \cite{RevModPhys.81.865}.

In particular, the entropy of entanglement is widely accepted as a measure for pure states of bipartite systems \cite{PhysRevA.56.R3319}, while measures such as the entanglement of formation \cite{PhysRevLett.80.2245}, entanglement distillation \cite{PhysRevA.54.3824, PhysRevLett.76.722, PhysRevLett.80.5239}, and related entropic quantities \cite{PhysRevLett.78.2275} have been introduced and validated for mixed bipartite states as well \cite{Adesso_2016}.

A rich literature also focuses on multipartite entanglement, where several approaches have been proposed. These include the classification of pure states into equivalence classes under local operations \cite{PhysRevA.62.062314, briegel_PRL86_910}, and the use of generalized entanglement measures such as the Schmidt measure or extensions of concurrence to mixed multipartite states \cite{PhysRevA.64.022306, PhysRevResearch.2.043062, PhysRevA.61.052306, PhysRevLett.93.230501}.
On top of these lines of work, the recent growth of quantum technologies has led some of the authors to use Entanglement Distance to study quantum graphs and quantum graph based structures \cite{VesperiniF24,DeSimoneF25,DeSimoneF25_1}, line of work also shared by other groups \cite{Gnatenko21,GnatenkoL22,Gnatenko23,Gnatenko24,Gnatenko26,gori2025multipartite}.
The Entanglement Distance is a measure derived from a geometric approach to quantum entanglement \cite{PhysRevA.101.042129, ScienticReports1-13, VesperiniB24}. It is important to emphasize that, already around three decades ago, entanglement was investigated from a geometric perspective. In particular, in a pioneering work, Shimony \cite{SHIMONY} introduced one of the earliest geometric definitions of the degree of entanglement for pure quantum states: the Geometric Entanglement Measure (GEM). In this framework, entanglement is interpreted as the squared Hilbert-space distance to the nearest separable state. Shortly thereafter, T. C. Wei and P. M. Goldbart \cite{PhysRevA.68.042307} reformulated and generalized Shimony’s idea in terms of overlaps rather than Euclidean distances, the measure was extended to mixed states via the convex roof construction and was proven to be an entanglement monotone.

The Entanglement Distance arises from a different geometric principle. Instead of measuring the distance to the set of separable states, ED is derived from the intrinsic Riemannian geometry of projective Hilbert space equipped with the Fubini–Study metric. Local unitary transformations generate orbits onto the projective Hilbert space. The elements of each orbit share the same degree of entanglement. Thus, ED is defined through the pullback of the Fubini–Study metric onto local unitary orbits.

Note that several other entanglement measures based on geometric correlations have been proposed in the literature, such as those based on the Bures distance \cite{PhysRevA.68.062309} or on the Hellinger geometric discord \cite{Marian_2015}. However, ED has the advantage of admitting a closed, explicit mathematical expression that does not require any minimization procedure, unlike these latter measures.
In addition, estimation-oriented approaches based on statistical distance concepts—such as the quantum Fisher information—have been proposed to characterize and quantify entanglement, particularly in metrological contexts \cite{PhysRevLett.72.3439, EPJD, PhysRevLett.102.100401, PhysRevA.85.022321, PhysRevA.85.022322, Tóth_2014, j.aop.2019.167995}.

In this work, we derive the Fubini–Study metric \cite{cmp1103908308, gibbons, BRODY200119}, which endows the manifold of multi-qubit quantum states with a Riemannian structure.
We then explore the deep relationship between this Riemannian structure—defined on the projective Hilbert space of the system—and the entanglement properties of the states it contains.
Thus, we derived a measure of entanglement, the entanglement distance (ED), denoted by $E$, a quantity preliminarily introduced in Ref. \cite{PhysRevA.101.042129} by some of the present authors and later studied further in Ref. \cite{VesperiniB24}.

In addition, we establish the remarkable properties of ED that render it a suitable measure of entanglement for multipartite quantum systems.

\section{What is quantum entanglement?}\label{secwis}
Quantum entanglement is a fundamental phenomenon in quantum physics whereby two or more subsystems exhibit correlations so strong that their physical properties cannot be described independently, even when the subsystems are spatially separated by large distances. More precisely, entanglement is the property of a composite quantum system such that a measurement performed on one subsystem—whose outcome is not determined a priori—affects the statistical outcomes of measurements performed on another subsystem, whose outcome too is not determined a priori, independently of the spatial separation between them.

\section{Pure-State Entanglement Distance}\label{secED}
Quantum mechanics can be regarded as an inherently geometric theory. From this perspective, a powerful geometrical framework is provided by the Riemannian metric structure defined on the manifold of quantum states. The Hilbert space is equipped with a Hermitian inner product, which naturally induces a notion of distance between state vectors. Let ${\cal H}$ denote the Hilbert space of a general quantum system. Given two nearby vectors in ${\cal H}$, $| \psi_1 \rangle$ and $| \psi_2 \rangle$, the scalar product $\langle \psi_1|\psi_2\rangle$, induces the norm $\Vert \Vert$ and, consequently, a (finite) distance between the two vectors, defined as
\begin{equation}
D(| \psi_1 \rangle,|\psi_2\rangle)=\Vert | \psi_1 \rangle - |\psi_2\rangle\Vert =
\langle \psi| \psi \rangle^{1/2} \, ,
\end{equation}
where $|\psi\rangle = |\psi_1\rangle -|\psi_2\rangle$. In the case of 
two normalized vectors $| \psi_1 \rangle$ and $ |\psi_2\rangle$, it results
\begin{equation}
D(| \psi_1 \rangle,|\psi_2\rangle)=  \left[ 2 \left( 1- \Re (\langle \psi_1  |\psi_2\rangle ) \right) \right]^{1/2} \, .
\end{equation}
Furthermore, the Hilbert space carries the structure of a differentiable manifold, so that it is always possible to introduce a local chart on ${\cal H}$ containing two nearby states. This, in turn, allows one to derive the metric tensor induced by the distance defined above.
Let $|\psi \rangle$ and $|\psi\rangle + |d\psi\rangle$ be two neighboring vectors. The squared (infinitesimal) distance between them is obtained by expanding the distance $D$ up to second order, yielding
\begin{equation}
d^2(|\psi\rangle + |d\psi\rangle,|\psi\rangle)=\langle d\psi  |d\psi\rangle \, .
\end{equation}
Thus, by means of a local chart, the normalized vectors in ${\cal H}$ smoothly depend on $N$-dimensional parameter $\xi\in \mathbb{R}^N$ and one has
\begin{equation}
|d\psi\rangle = \sum_\mu |\partial_\mu \psi(\xi) \rangle d\xi^\mu \, ,
\end{equation}
where with $\partial_\mu \psi$ we mean ${\partial \psi}/{\partial \xi^\mu}$. 
Thus one has
\begin{equation}
d^2(|\psi\rangle + |d\psi\rangle,|\psi\rangle) = \sum_{\mu \nu}\langle \partial_\mu \psi|\partial_\nu \psi \rangle d\xi^\nu  d\xi^\mu \, .
\end{equation}
Although the matrix elements $\langle\partial_\mu \psi | \partial_\nu \psi \rangle$ may appear to define the components of a Riemannian metric tensor on ${\cal H}$, they do not have a direct physical interpretation as a distance between quantum states. In fact, the Hilbert space provides a redundant description of quantum states: physical states are associated with rays in Hilbert space, and two normalized kets that differ only by a phase factor $e^{i\alpha}$ represent the same quantum state.
Therefore, consistency requires that the distance between$\lvert \psi_1 \rangle$ and $\lvert \psi_2 \rangle$ be the same as the distance between $e^{i\alpha}\lvert \psi_1 \rangle$ and $e^{i\beta}\lvert \psi_2 \rangle$, for any real $\alpha$ and $\beta$.
By introducing a local chart, this requirement can be formulated in a precise mathematical framework: an appropriate metric tensor on the space of states must be invariant under the gauge transformation $|\psi(\xi)\rangle \to e^{i\alpha(\xi)} |\psi(\xi)\rangle$.
This requirement is satisfied by the Fubini–Study metric, which defines the (squared) distance between two neighbouring rays.
\begin{equation}
d^2_{FS}(| \psi\rangle + |d \psi\rangle,|\psi\rangle)=\langle d\psi |d\psi\rangle -\langle \psi |d\psi\rangle \langle d\psi|\psi\rangle \, ,
\label{F-S-metric}
\end{equation}
from which one derives the metric tensor
\begin{equation}
g_{\mu \nu} = \langle \partial_\mu \psi|\partial_\nu \psi \rangle -
\langle \partial_\mu \psi| \psi \rangle \langle \psi|\partial_\nu \psi \rangle \, .
\label{mtFS}
\end{equation}
The Fubini–Study metric \eqref{F-S-metric} is therefore defined on the finite projective Hilbert space ${\cal PH}$ \cite{cmp1103908308, gibbons}, namely on the set of equivalence classes of nonzero vectors 
$|\psi\rangle \in {\cal H}$ under the equivalence relation $\thicksim_p $ on ${\cal H}$, defined by $|\psi\rangle  \thicksim_p |\phi\rangle$ if and only if $|\psi\rangle = \alpha |\phi\rangle$ for some $\alpha \in \mathbb{C}$, with $\alpha \neq 0$.

It is worth remarking that one can define the square of the (finite) distance between two rays $[|\phi_1\rangle]_p,[|\phi_2\rangle]_p \in {\cal PH}$, associated with the normalized states $e^{i\alpha_1}|\phi_1\rangle,e^{i\alpha_2}|\phi_2\rangle$, respectively, as follows \footnote{Note that in the literature two distinct definitions of the non-infinitesimal Fubini--Study metric can be found. On the Bloch sphere, these correspond, up to a scale factor, to {\it (i)} the straight-line (chordal) distance between two points on the sphere, and {\it (ii)} the geodesic distance along the spherical surface. In the present work, we adopt definition {\it (i)}. The two definitions coincide in the infinitesimal limit.}
\begin{equation}
D^2_{FS}(|\phi_1\rangle,|\phi_2\rangle) = (1-|\langle\phi_1|\phi_2 \rangle|^2) \, .
\label{fFS}
\end{equation}
One can easily verify that the latter distance induces the metric tensor \eqref{mtFS}.
In fact, by expanding $|\phi_1\rangle$ up to second order as 
\begin{equation}
|\phi_1(\xi)\rangle=
|\psi\rangle + \sum_\mu |\partial_\mu \psi \rangle  d\xi^\mu + \frac{1}{2}\sum_{\mu\nu} |\partial^{2}_{\mu\nu} \psi \rangle d\xi^\mu d\xi^\nu \, ,
\end{equation}
and setting $|\phi_2\rangle =|\psi\rangle$, from  Eq. \eqref{fFS} one gets
\begin{equation}
D^2_{FS}(|\phi_1\rangle ,|\phi_2\rangle) =\sum_{\mu\nu} g_{\mu\nu} d\xi^\mu d\xi^\nu \, ,
\label{fFSdiff}
\end{equation}
where $g_{\mu\nu}$ is the one of Eq. \eqref{mtFS}.

To investigate the deep connection between the Riemannian metric structure associated with the projective Hilbert space and the entanglement properties of the states defined on this space, we equip the projective Hilbert space with a metric derived from the Fubini–Study metric, from which we obtain a meaningful definition of an entanglement measure.
We consider the case of the Hilbert space
${\cal H } = {\cal H}^0 \otimes {\cal H}^1 \cdots {\cal H}^{M-1}$ tensor product of $M$ qubits Hilbert spaces.

The entanglement measure is invariant under local unitary (LU) transformations. Therefore, given $[|\phi\rangle]_p,[|\psi\rangle]_p \in {\cal PH}$ and their associated normalized representatives $|\phi\rangle,|\psi\rangle \in {\cal H}$, we introduce the following equivalence relation on the projective Hilbert space
\begin{equation}
[|\phi\rangle]_p \thicksim [|\psi\rangle]_p \, , \quad \rm {iff} \ |\phi\rangle = e^{i\alpha} \prod^{M-1}_{\mu=0} U^\mu |\psi\rangle \, , 
\end{equation}
where, for $\mu=0,\ldots,M-1$, each operator $U^\mu$ is an arbitrary $SU(2)$ LU operator that operates on the $\mu$th qubit and $\alpha \in \mathbb{R}$. With this equivalence relation, one derives the quotient set ${\cal P H}/\thicksim $. Thus, the entanglement measure $E$ has to be a function $E: {\cal PH}/\thicksim \to \mathbb{R}^+$, that is a function of the equivalence classes of ${\cal P H}$ by $\thicksim$, that is
\begin{equation}
[|\psi\rangle] = \left\lbrace |\phi\rangle \in {\cal P H} | \  |\phi\rangle \thicksim |\psi \rangle  \right\rbrace\, .
\label{Us}
\end{equation}
Following Ref. \cite{PhysRevA.101.042129}, we derive an entanglement measure from a distance inspired from the Fubini-Study one. For each normalized ket $|\psi\rangle\in{\cal H}$  we consider
\begin{equation}
\left\lbrace |U,\psi\rangle = \prod^{M-1}_{\mu=0} U^\mu |\psi\rangle \right\rbrace\, ,
\label{Upsi}
\end{equation}
the set of all the vectors derived from $|\psi\rangle$ under the action of LU operators,
where, for $\mu=0,\ldots,M-1$, each operator $U^\mu$ is an arbitrary $SU(2)$ LU operator that operates on the $\mu$th qubit. 
Note that all kets in \eqref{Upsi} have the same degree of entanglement.
For each vector $|U,\psi\rangle $ in \eqref{Upsi}, we introduce a local chart in a neighborhood of that vector by means of the unitary operator
$
e^{-i\sum_{\mu=0}^{M-1} \bm{\sigma}_{\bf n}^\mu\xi^\mu } \, ,
$
which depends on real parameters $\xi^\mu$, where ${\bf n}^\mu$ 
are fixed unit vectors. With this construction, the point $\xi^\mu =0$ for $\mu=0,\ldots,M-1$ corresponds to the vector $|U,\psi\rangle $.
Here and in the following, we use the notation $\bm{\sigma}_{\bf n}^\mu = {\bf n}^\mu \cdot \bm{\sigma}^\mu$.
Furthermore, for $\mu=0,\ldots,M-1$, we denote by $\sigma^\mu_1$, $\sigma^\mu_2$, and $\sigma^\mu_3$ the three Pauli matrices acting on the $\mu$-th qubit, where the index $\mu$ labels the spins.
We consider an infinitesimal variation of ket $|U,\psi\rangle $ given by
\begin{equation}
|dU,\psi\rangle = \sum^{M-1}_{\mu=0} d\tilde{U}^\mu |U,\psi\rangle \, ,
\label{dUs}
\end{equation}
where 
\begin{equation}
d\tilde{U}^\mu = -i 
 \bm{\sigma}_{\bf n}^\mu
d \xi^\mu 
\label{dU}
\end{equation}
rotates the $\mu$th qubit by an infinitesimal angle $2 d\xi^\mu$  around
the unitary vector ${\bf n}^\mu $.

By substituting $|U,\psi\rangle$ and $|dU,\psi\rangle $ in Eq. \eqref{F-S-metric}, in place of $|\psi\rangle$ and $|d\psi\rangle$, respectively, we get 
\begin{equation}
    d^2_{_{FS}} (|U,\psi\rangle +|dU,\psi\rangle,|U,\psi\rangle ) =
    \sum_{\mu \nu }g_{\mu \nu} (|\psi\rangle,{\bf v}) d\xi^\mu d\xi^\nu \, ,
\end{equation}
where, the corresponding \textit{projective Fubini-Study metric tensor} is
\begin{equation}
    g_{\mu \nu} (|\psi\rangle,{\bf v}) =
\langle \psi | \bm{\sigma}_{\bf v}^\mu
\bm{\sigma}_{\bf v}^\nu|\psi\rangle 
-\langle \psi | \bm{\sigma}_{\bf v}^\mu
|\psi\rangle
\langle \psi |
\bm{\sigma}_{\bf v}^\nu|\psi\rangle \, ,
\label{gmunu}
\end{equation}
${\bf v} = ({\bf v}^0,\ldots,{\bf v}^{M-1})$ and
the unit vectors ${\bf v}^\mu$, $\mu=0,\dots,M-1$, are derived by a rotation of the original
ones of Eq. \eqref{dU}, according to $\bm{\sigma}^\nu_{\bf v} =
U^{\nu \dagger}  \bm{\sigma}^\nu_{\bf n}
U^{\nu}$,
where there is no summation on the index $\nu$.
Clearly, for a given state $|\psi\rangle$, the metric tensor $g_{\mu \nu} (|\psi\rangle,{\bf v})$ is not invariant under rotations of the unit vectors  ${\bf v}^\mu$. In order to construct a quantity that is invariant under such rotations, we define the entanglement measure associated with the equivalence class $[|\psi\rangle]$ as the infimum of the trace of $g_{\mu \nu} (|\psi\rangle,{\bf v})$ over all possible orientations of the unit vectors ${\bf v}^\mu$. In formulae, we define the ED as
\begin{equation}
E(|\psi\rangle) =\inf_{ \{{\bf v}^\nu\}_\nu} \tr (g(|\psi\rangle,{\bf v})) \, ,
\label{emeasure}
\end{equation}
where $\tr$ is the trace operator and where the $\inf$ is taken over all possible orientations of the unit vectors $ {\bf v}^\nu$ ($\nu = 0, \dots , M-1$).
We emphasize that, in general, inspection of the block structure of $g(|\psi\rangle)$ provides valuable information about $k$-separability. Consider a choice of unit vectors $ {\bf v}^\nu$ that induces a metric $g(|\psi\rangle,{\bf v})$ which, up to a permutation of the qubit indices, is block-diagonal.
In a previous work by one of the authors \cite{VESPERINI2023169406}, it was shown that $n\geq p\geq k$, where $n$ denotes the number of such blocks, $p$ the persistency of entanglement, and $k$ the degree of separability.
In particular, this result implies that if there exists a choice of unit vectors such that $g(|\psi\rangle,{\bf v})$ is irreducible for any permutation of its indices (i.e., $n=1$), then the state $|\psi\rangle$ is genuinely multipartite entangled (i.e., $k=1$).

From Eq. \eqref{gmunu} we derive
\begin{equation}
    \tr [ g(|\psi\rangle,{\bf v}) ]= \sum^{M-1}_{\mu=0} \left[1 - ({\bf v}^\mu\cdot \langle \psi |\bm{\sigma}^\mu|\psi\rangle )^2\right] \, ,
\label{ed1}
\end{equation}
that shows that the unit vectors
\begin{equation}
\tilde{\bf v}^\mu = \pm \langle \psi |\bm{\sigma}^\mu|\psi\rangle/\Vert \langle \psi |\bm{\sigma}^\mu|\psi\rangle \Vert \, ,
    \label{vtildes}
\end{equation}
provide the $\inf$ of $\tr(g)$.
Therefore, we obtain the following directly computable formula for the ED
\begin{equation}
    E(|\psi\rangle) =M - \sum^{M-1}_{\mu =0} \Vert \langle \psi |\bm{\sigma}^\mu|\psi\rangle\Vert^2 \, .
    \label{measureMqpure}
\end{equation}
Note that the latter equation can be seen as the sum of the $M$ single-qubit EDs \begin{equation}\label{measureMqpure_1qubit}
E_\mu(|\psi\rangle)=1 - \Vert \langle \psi |\bm{\sigma}^\mu|\psi\rangle\Vert^2.
\end{equation}
$E_\mu(|\psi\rangle)$ is a measure of bipartite entanglement of $\mu$ with the rest of the system. Note that Eq. \eqref{measureMqpure} also has the meaning of a quantum correlation measure \cite{ScienticReports1-13}.

The $\inf$ operation renders the measure in \eqref{emeasure} independent of the choice of the operators $U^\mu$. Consequently, its numerical value is associated with the equivalence class \eqref{Us} and does not depend on the particular representative chosen within the class. This is a necessary requirement for a well-defined entanglement measure \cite{PhysRevLett.78.2275}.

Remarkably, the entanglement measure can be derived from a minimum-distance principle when formulated within the framework of the Riemannian geometry of the projective Hilbert space. Indeed, according to Eq. \eqref{fFS}, the squared distance between the rays associated with the unit vectors $|\phi\rangle$ and $|\phi^\mu({\bf v}^\mu)\rangle\equiv\bm{\sigma}_{\bf v}^\mu |\phi\rangle$ is
\begin{equation}
D^2_{FS}(|\phi\rangle,|\phi^\mu({\bf v}^\mu)\rangle) =
1-|\langle \phi | \phi^\mu({\bf v}^\mu)\rangle|^2 \, .
\end{equation}
We name  ${\bf v}^\mu$-conjugate of $|\phi\rangle$ the states $|\phi^\mu({\bf v}^\mu)\rangle$, for $\mu=0,\ldots M-1$.
Therefore 
\begin{equation}
E(|\phi\rangle) = \inf_{ \{{\bf v}^\nu\}_\nu} \sum^{M-1}_{\mu=0}
D^2_{FS}(|\phi\rangle,|\phi^\mu({\bf v}^\mu)\rangle) \, .
\end{equation}
This shows that the minimum of the sum of the squared (finite) distances between a state $|\phi\rangle$ and all the states obtained by the action of the operators $\bm{\sigma}_{\bf v}^\mu$, upon varying the vectors ${\bf v}^\mu$, is bounded from below by the entanglement measure $E(|\phi\rangle)$.
For fully separable states, this minimum distance vanishes, whereas for maximally entangled states it reaches, at best, the value $M$.
Therefore, within this geometric framework, entanglement can be interpreted as an obstruction to minimizing the sum of the squared distances between a state $|\phi\rangle$ and all its ${\bf v}^\mu$-conjugate states.

\subsection{Properties of Entanglement Distance}\label{secpem}

The ED is an entanglement monotone \cite{vidal_2000,PhysRevLett.78.2275}  in the sense it fulfils the following conditions:
\begin{enumerate}[label=\roman*.]

\item  $E(|\psi\rangle) \geq 0$, and $E(|\psi\rangle)= 0$ if and only if $|\psi\rangle$ is fully separable;

\item $E$ is invariant under LU transformation;

\item \label{terza} $E$  doesn't increase under local operation and classical communications (LOCC);

\item $E$ is additive for tensor products. 

\end{enumerate}

In fact:

\begin{enumerate}[label=\roman*.]

\item From \eqref{measureMqpure_1qubit} it follows that $0\leq E_\mu(|\psi\rangle)\leq 1$, since $ 0\leq \Vert \langle \psi |\bm{\sigma}^\mu|\psi\rangle \Vert^2 \leq 1$. Therefore, $E(|\psi\rangle)= 0$ implies $E_\mu(|\psi\rangle)= 0$ for each $\mu$.
The reduced density matrix of the $\mu$-th subsystem, $\rho^\mu = \tr_{\nu\neq \mu} [\ket{\psi} \bra{\psi}]$, obtained by tracing over the degrees of freedom of the remaining subsystems, can be written as $\rho^\mu =(\mathbb{I} +  \bm{\sigma}_{\bf v}^\mu )/2$.
Hence, $E_\mu(|\psi\rangle)= 0$ implies $\Vert \langle \psi |\bm{\sigma}^\mu|\psi\rangle \Vert^2 = 1$. Since $\Vert \langle \psi |\bm{\sigma}^\mu|\psi\rangle \Vert^2 =\Vert {\bf v}^\mu \vert^2 $,
it follows that $\Vert {\bf v}^\mu \vert^2 = 1$ for each $\mu$. This condition is satisfied if and only if the state $\ket{\psi}$ is fully separable.

\item For a given LU operator $U$, which has the form $U=\otimes_\nu U^\nu$ with $U^\nu$ a unitary operator acting on the $\mu$-th party, one finds that $\Vert \langle \psi |U^{\dagger}\bm{\sigma}^\mu U|\psi\rangle \Vert^2 =  \Vert  \langle \psi |\bm{\sigma}^\mu|\psi\rangle  \Vert^2$.  This proves the statement.

\item Let us suppose that a local measurement is performed on a single qubit $\mu$; without loss of generality, we may take $\mu=0$. If $\ket{\psi}$ denotes the normalized state vector before the measurement, then $\ket{\tilde{\psi}}$ denotes the normalized state vector after the measurement, corresponding to the outcome $+1$ for the qubit $\mu=0$ along the direction ${\bf v}$. We denote the corresponding eigenstate of the measured qubit by $\ket{{\bf v}}^0$.
The associated outcome probability is 
$
P_{{\bf v}} = |\bra{\psi} \ket{\tilde{\psi}}|^2
$.
The post-measurement state vector is given by
\begin{equation}
    \ket{\tilde{\psi}}=\frac{\ket{{\bf v}}^0\bra{{\bf v}}^0\ket{\psi}}{\Vert \ket{{\bf v}}^0\bra{{\bf v}}^0\ket{\psi} \Vert} \, .
\end{equation}
It results
\begin{equation}
P_{{\bf v}} = \Vert {\ket{{\bf v}}^0} \bra{{\bf v}}^0\ket{\psi} \Vert^2 \leq 1 \, .
\end{equation}
The reduced density matrix of the $\nu$-th qubit is given by
\begin{equation}
    \tilde{\rho}^\nu = \tr_{\mu\neq \nu} [\ket{\tilde{\psi}}\bra{\tilde{\psi}}] \, .
\end{equation}    
For $\nu=0$, one obtains
\begin{equation}  
    \tilde{\rho}^0 =
    \ket{{\bf v}}^0\bra{{\bf v}}^0 \, .
\end{equation}
Therefore, from \eqref{measureMqpure_1qubit} we find
\begin{equation}
    E_0(\ket{\tilde{\psi}})= 0 \leq E_0(\ket{{\psi}})\, .
\end{equation}
The two-qubit reduced density matrix of the $0$-th and the $\nu$-th qubits, with $\nu\neq 0$, is given by
\begin{equation}
    \tilde{\rho}^{0 , \nu} = \tr_{\mu\neq 0,\nu} [\ket{\tilde{\psi}}\bra{\tilde{\psi}}] \, .
\end{equation}

By direct calculation, one can verify that
\begin{equation}
\begin{split}
    \tr[\sigma^\nu_i \tilde{\rho}^\nu] & =
    \\
&    \dfrac{1}{P_{{\bf v}}}
    \sum_{i_0,i_\nu,j_0,j_\nu =0,1}  \bra{\bf v}^0 \bra{i_\nu}^\nu 
    \tilde{\rho}^{0 , \nu} \ket{\bf v}^0 \ket{j_\nu}^\nu \geq
    \\
    &
    \sum_{i_0,i_\nu,j_0,j_\nu =0,1}  \bra{\bf v}^0 \bra{i_\nu}^\nu 
    \tilde{\rho}^{0 , \nu} \ket{\bf v}^0 \ket{j_\nu}^\nu \geq
    \tr[\sigma^\nu_i {\rho}^\nu] \, ,
\end{split}
\end{equation}
this proves that, for $\nu\neq 0$,
\begin{equation}
    E_\nu (\ket{\tilde{\psi}}) \leq E_\nu(\ket{{\psi}}) \, .
\end{equation}
This completes the proof of Claim~\ref{terza}.

\item In the case of a state $\ket{\psi}$ product of two states $\ket{\psi}=\ket{\psi_1}\otimes \ket{\psi_2}$, the metric tensor $g(\ket{\psi},{\bf v})$ can be set as diagonal blocks
\begin{equation}
g(\ket{\psi},{\bf v}) =
\left(
\begin{array}{c c}
g_1(\ket{\psi_1},{\bf v}) & 0 \\
0 & g_2(\ket{\psi_2},{\bf v})
\end{array}
\right) \, ,
\end{equation}
and one has
\begin{equation}
E(\ket{\psi}) = E(\ket{\psi_1}) + E(\ket{\psi_2}) \, .
\end{equation}
The generalization to multiple tensor products follows straightforwardly.
\end{enumerate}

\subsection{Comparison between the concurrence and the entanglement distance.}

Let us consider a general $M=2$ qubits normalized pure-state
\begin{equation}
    |\psi\rangle = \sum^3_{j=0} w_j |j\rangle \, ,
    \label{2btstate}
\end{equation}
such that $  \sum^3_{j=0} |w_j|^2 = 1$.
Concurrence, for pure states \eqref{2btstate}, is defined as \cite{10.5555/2011326.2011329, sym15112089}
\begin{equation}
    C(|\psi\rangle )= |\langle \psi|\psi^\dagger\rangle| \, ,
    \label{concupsi}
\end{equation}
where $    |\psi^\dagger \rangle = (\sigma^0_2 \otimes \sigma^1_2)  \sum^3_{j=0} w^*_j |j\rangle $.
By direct computations one gets \cite{10.5555/2011326.2011329}
\begin{equation}
    C(|\psi\rangle) = 2 |w_0 w_3-w_1 w_2| \, .
    \label{concupsi1}
\end{equation}
For the same case ED \eqref{measureMqpure} can be written explicitly in terms of the amplitudes $w_i$ as 
\begin{equation}\label{ED_2qubits}
    E(|\psi\rangle )= 8[|w_0|^2 |w_3|^2 + |w_1|^2 |w_2|^2-\! w^*_0w^*_3 w_1 w_2-\! w_0w_3 w^*_1 w^*_2]
    \,,
\end{equation}
which can be written as a monotone function of the concurrence
\begin{equation}
    E(|\psi\rangle) = 2[C(|\psi\rangle)]^2 \, .
    \label{EMvsCpure}
\end{equation}
This proves that the concurrence for pure states is a special case of ED, valid for the case of two qubits.

\subsection{Comparison between the entanglement entropy and the entanglement distance.}

In the special case of pure two-qubit states, the entropy of entanglement, $E_S(|\Psi\rangle)$, can be expressed explicitly as a function of the entanglement distance $E_D(|\Psi\rangle)$. 

Indeed, the relation between the two-qubit concurrence and the entanglement entropy (for pure states) is well known and monotonic. Thus, using \eqref{EMvsCpure}, one finds
\begin{equation}
E_S(|\Psi\rangle)
=
k_B\,F\!\left(\frac{1+\sqrt{1-\frac{E_D(|\Psi\rangle)}{2}}}{2}\right),
\label{eq:ES_of_ED}
\end{equation}
where
\begin{equation}
F(x) = -x\ln x - (1-x)\ln(1-x).
\end{equation}
This relation follows directly from \eqref{EMvsCpure}, see, for instance, Ref.~\cite{PhysRevLett.80.2245}, Eq.~(9).

\section{Example: calculation of the Entanglement Distance}

As an illustrative example of the calculation of the Entanglement Distance, we consider the following normalized three-qubit state:
\begin{equation}
\ket{\psi(\theta,\phi)} = 
c_2 \big( c_1 \ket{00} + s_1 \ket{11} \big)\ket{0} 
+ s_2 \big( c_1 \ket{01} - s_1 \ket{11} \big)\ket{1} \, ,
\label{stateex}
\end{equation}
where
\begin{equation}
\begin{split}
c_1 &:= \cos(\theta) \, , \\
s_1 &:= \sin(\theta) \, , \\
c_2 &:= \cos(\phi) \, , \\
s_2 &:= \sin(\phi) \, ,
\end{split}
\end{equation}
with $\theta,\phi \in [0,\pi/2]$.

One can verify that for $\phi = 0$ and $\theta \neq 0,\pi/2$, the state 
$\ket{\psi(\theta,0)}$ is biseparable, being the product of a two-qubit state 
and a single-qubit state. Moreover, the states $\ket{\psi(0,0)}$ and 
$\ket{\psi(\pi/2,0)}$ are fully separable. For $\phi = \pi/2$, the states 
$\ket{\psi(\theta,\pi/2)}$ are fully separable for all values of $\theta$.
For $\theta = \pi/2$, the states 
$\ket{\psi(\pi/2,\phi)}$ are fully separable for all values of $\phi$.

By direct calculation, one obtains the following expectation values of the 
Pauli matrices for the first qubit:
\begin{equation}
\begin{split}
\bra{\psi(\theta,\phi)} \sigma^{(1)}_1 \ket{\psi(\theta,\phi)} 
&= - 2 c_1 s_1 s_2^2 \, , \\
\bra{\psi(\theta,\phi)} \sigma^{(1)}_2 \ket{\psi(\theta,\phi)} 
&= 0 \, , \\ 
\bra{\psi(\theta,\phi)} \sigma^{(1)}_3 \ket{\psi(\theta,\phi)} 
&= c_1^2 - s_1^2 \, ,
\end{split}
\end{equation}
for the second qubit:
\begin{equation}
\begin{split}
\bra{\psi(\theta,\phi)} \sigma^{(2)}_1 \ket{\psi(\theta,\phi)} 
&= 0 \, , \\
\bra{\psi(\theta,\phi)} \sigma^{(2)}_2 \ket{\psi(\theta,\phi)} 
&= 0 \, , \\ 
\bra{\psi(\theta,\phi)} \sigma^{(2)}_3 \ket{\psi(\theta,\phi)} 
&= (c_1^2 - s_1^2) c_2^2 - s_2^2 \, ,
\end{split}
\end{equation}
and for the third qubit:
\begin{equation}
\begin{split}
\bra{\psi(\theta,\phi)} \sigma^{(3)}_1 \ket{\psi(\theta,\phi)} 
&= - 2 s_1^2 s_2 c_2 \, , \\
\bra{\psi(\theta,\phi)} \sigma^{(3)}_2 \ket{\psi(\theta,\phi)} 
&= 0 \, , \\ 
\bra{\psi(\theta,\phi)} \sigma^{(3)}_3 \ket{\psi(\theta,\phi)} 
&= c_2^2 - s_2^2 \, .
\end{split}
\end{equation}

\begin{figure}[ht!]
    \centering
    \includegraphics[width=\linewidth]{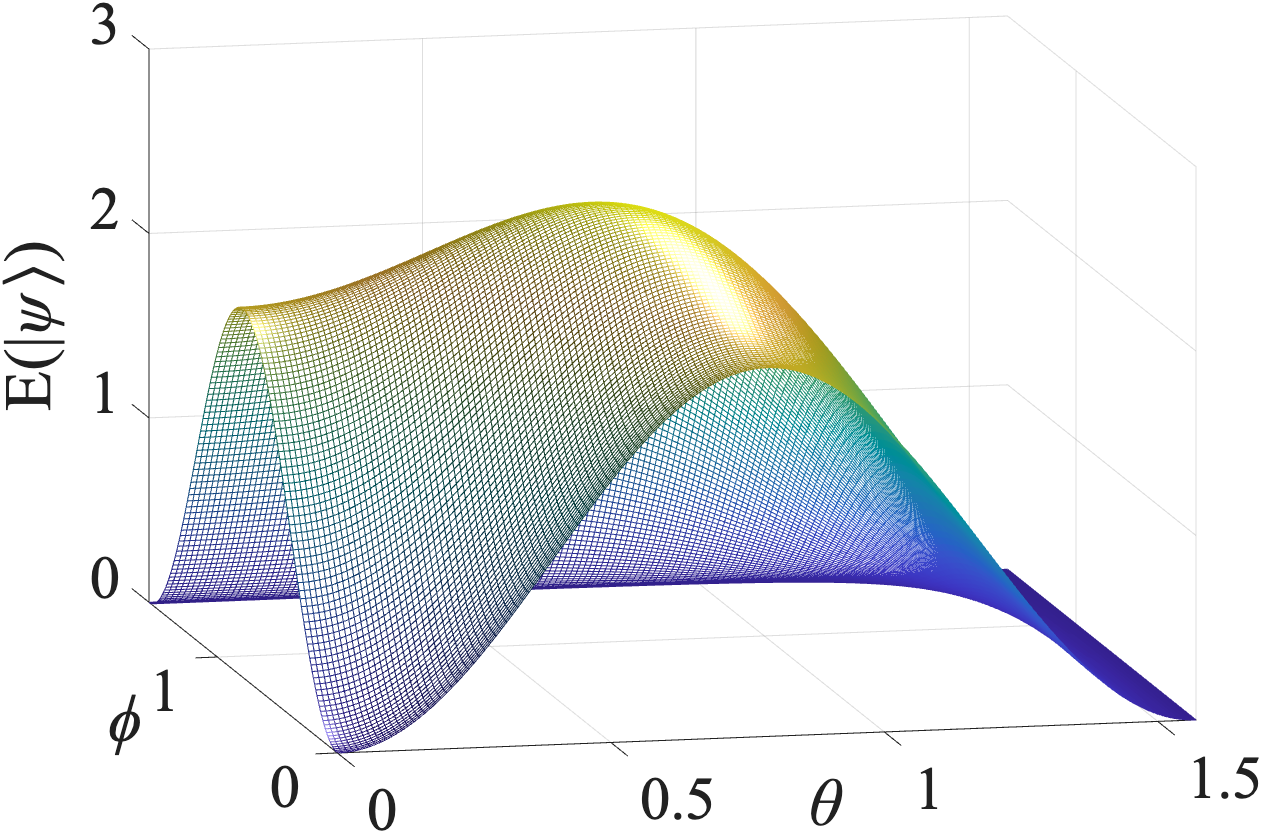}
    \caption{This figure reports the three-dimensional plot of the ED $E(\ket{\psi})$ as a function of  $\theta$ and $\phi$ for the states \eqref{stateex}.}
    \label{fig}
\end{figure}

Finally, by Eq.~\eqref{measureMqpure}, one obtains
\begin{equation}
\begin{split}
E(\ket{\psi(\theta,\phi)}) 
= 3 - \frac{1}{16} \Big[ 
&31 - 8 \cos(2\theta) + 9 \cos(4\theta) \\
&- 4 \cos^2(\theta) \cos(4\phi)\big(\cos(2\theta)-5\big) \\
&- 16 \cos(2\phi)\sin^2(2\theta)
\Big] \, .
\end{split}
\end{equation}
Figure \ref{fig} shows the $3D$ plot of $E(\ket{\psi(\theta,\phi)})$ as a function
of $\theta,\phi\in[0,\pi/2]$.
The entanglement measure $E(\ket{\psi(\theta,\phi)})$ correctly captures all the  entanglement properties expected for the state~\eqref{stateex}.

\section{Concluding remarks}\label{seccr}

By framing entanglement through the lens of the Fubini-Study metric, we have demonstrated that the Riemannian geometry of the projective Hilbert space offers a deep, intuitive foundation for quantum correlations. Within this framework, the entanglement distance (ED) emerges as more than just a formal construction; it represents a physical "obstruction" that prevents a state from being locally minimized toward a separable form.

This geometric intuition is rigorously validated in the bipartite regime. For pure states of two qubits, we have shown that the ED is an exact monotone of both concurrence and the entropy of entanglement. Such results provide a bridge between differential geometry and established information-theoretic definitions, confirming that our geometric measure remains consistent with traditional benchmarks.

However, the power of the geometric approach lies in its potential to describe the more complex landscape of multipartite systems where standard measures often falter. While the transition to mixed states and the exploration of correlations beyond entanglement remain open challenges, the results presented here suggest that geometry provides a uniquely transparent language for these problems. Continuing to peel back the geometric layers of the projective Hilbert space will likely yield further insights into the structural role of entanglement in complex quantum systems.

%
%

%
%
%

\begin{acknowledgments}

We acknowledge the support of the Research Support Plan 2022 – Call for applications for funding allocation to research projects curiosity-driven (F CUR) – Project "Entanglement Protection of Qubits’ Dynamics in a Cavity" – EPQDC and the support from the Italian National Group of Mathematical Physics (GNFM-INdAM). R. F. would like to acknowledge INFN Pisa for the financial support for this activity.
\end{acknowledgments}

\bibliography{sn-bibliography}

\end{document}